\documentclass[preprint,12pt]{elsarticle}

\usepackage{graphics}
\usepackage{graphicx}
\usepackage{wrapfig}
\usepackage{amssymb}
\def\eq#1{{(\ref{#1})}}
\def\fig#1{{Fig.~\ref{#1}}}

\def\dm{{\partial}_{\mu}}
\def\eps{\epsilon^{\mu \nu \lambda \sigma}}
\def\d{\partial}

\newcommand{\beq}{\begin{equation}}
\newcommand{\eeq}{\end{equation}}
\newcommand{\ben}{\begin{eqnarray*}}
\newcommand{\een}{\end{eqnarray*}}
\journal{Annals of Physics}

\begin{document}

\begin{frontmatter}



\title{Topologically induced local ${\cal P}$ and ${\cal CP}$ violation\\ in ${\rm QCD \times QED}$ }


\author{Dmitri E. Kharzeev}

\address{Physics Department, Brookhaven National Laboratory, Upton, NY 11973-5000, USA\\
and \\
Department of Physics, Yale University, New Haven, CT 06520-8120, USA}

\begin{abstract}
The existence of topological solutions and axial anomaly open a possibility of 
$\cal{P}$ and $\cal{CP}$ violation in QCD. For a reason that has not yet been established conclusively, this possibility is not realized in strong interactions -- the experimental data indicate that a global  $\cal{P}$ and $\cal{CP}$ violation in QCD is absent. Nevertheless, the fluctuations of topological charge in QCD vacuum, although not observable directly, are expected to play an important r\^{o}le in the breaking of $U_A(1)$ symmetry and in the mass spectrum and other properties of hadrons. Moreover,  
in the presence of very intense external electromagnetic fields topological solutions of QCD can induce {\it local} ${\cal P}-$ and ${\cal CP}-$ odd effects in the 
$SU_c(3)\times U_{em}(1)$ gauge theory that can be observed in experiment directly. Here I show how these local parity-violating phenomena can be described by using the Maxwell-Chern-Simons, or axion, electrodynamics as an effective theory. Local ${\cal P}-$ and ${\cal CP}-$ violation in hot QCD matter can be observed in experiment through the "chiral magnetic effect" -- the separation of electric charge along the axis of magnetic field. Very recently, STAR Collaboration presented an observation of the electric charge asymmetry with respect to reaction plane in relativistic heavy ion collisions at RHIC. %Topologically induced local parity violation can play an important r^{o}le in the Early Universe, e.g. by providing a seed for the magnetic helicity. The related phenomena can also be modeled in condensed matter systems, including graphene.

\end{abstract}


\end{frontmatter}

\newpage
\section{Introduction}
\label{}

At present Quantum Chromo-Dynamics (QCD) is firmly established as the theory of strong interactions. 
At short distances the QCD running coupling constant is small due to the asymptotic freedom \cite{Gross:1973id,Politzer:1973fx}; because of this, the 
amplitudes of hard parton scattering can be reliably evaluated in perturbation theory. Moreover at weak coupling 
one expects to find a r\^{o}le for the classical solutions of QCD equations of motion as the quantum corrections to these solutions are under control. At strong coupling, we still lack a reliable theoretical method for performing analytical calculations in QCD. Nevertheless the results of numerical lattice calculations as well as analytical results in related theories suggest that some of the phenomena induced by the presence of classical solutions persist even when the coupling constant grows large. 

The non-Abelian nature of QCD has very important implications for the structure of the classical solutions. For example, the classical Yang-Mills equations (non-Abelian analogs of Maxwell equations) possess non-trivial vacuum solutions -- instantons -- 
corresponding to the mapping 
of the $SU(2)$ subgroup of the gauge group $SU(3)$ onto the group of three-dimensional rotations $S_3$ \cite{Belavin:1975fg}.    
Instantons thus couple rotations in space to rotations in the space of color. In the presence of fermions, this property of instantons causes non-conservation of chirality that would otherwise be forbidden by the conservation of angular momentum -- but the coupling of angular momentum to color allows to compensate the flip of spin by a rotation in color space.   
In Euclidean space-time, instantons are static localized objects. It is well known that the amplitude $A_{tun}$ of quantum tunneling in Minkowski space is determined in the quasi--classical approximation by the action $S_{cl}$ of the classical Euclidean solution, $A_{tun} \sim \exp(-S_{cl}/\hbar)$. Likewise, in Minkowski space-time instantons describe the tunneling transitions between the states with different topological "winding" numbers  $\nu$ of the $SU(2) \leftrightarrow S_3$ mapping \cite{tHooft,'tHooft:1976fv,Jackiw:1976pf,Callan:1976je}. These topological numbers can be represented as the four-dimensional space-time integrals over the divergence of Chern-Simons current \cite{Chern:1974ft}.

Topological fluctuations are believed to play an important r\^{o}le in the structure of QCD vacuum and in the properties of hadrons (for a review, see \cite{Schafer:1996wv}). They also put in doubt the ${\cal P}$ and ${\cal CP}$ invariances of QCD ("the strong ${\cal CP}$ problem"). However until now all of the evidence for the topological effects in QCD from experiment, however convincing, has been indirect. In this paper I will summarize the arguments for the possibility to observe the topological effects in QCD {\it directly} in the presence of very intense external electromagnetic fields. In particular, the coupling of topological gluon field configurations to electromagnetism induced by the axial anomaly leads to the separation of electric charges in the presence of a strong external magnetic field \cite{Kharzeev:2004ey} ("the chiral magnetic effect" \cite{Kharzeev:2007tn,Kharzeev:2007jp,Fukushima:2008xe,Kharzeev:2009pj}). The evidence for the chiral magnetic effect has been found recently from the numerical lattice QCD calculations \cite{Buividovich:2009wi}. The magnetic fields of the required  strength can be created in heavy ion collisions \cite{Kharzeev:2007jp,Skokov:2009qp}. The direction of the produced magnetic field is orthogonal to the reaction plane of the collision. Very recently, STAR Collaboration presented an observation of the electric charge asymmetry with respect to reaction plane in relativistic heavy ion collisions at RHIC \cite{Abelev:2009uh,Abelev:2009tx}.

\section{The strong ${\cal CP}$ problem} 
\vskip0.3cm

 Let us begin with a brief introduction to the strong ${\cal CP}$ problem.  
Strong interactions within the Standard Model are described by Quantum Chromo-Dynamics, with the Lagrangian dictated 
by the $SU(3)$ color gauge invariance: 
\beq \label{qcd}
{\cal L} = -{1 \over 4} G^{\mu\nu}_{\alpha}G_{\alpha \mu\nu}  + \sum_f \bar{\psi}_f \left[ i \gamma^{\mu} 
(\partial_{\mu} - i g A_{\alpha \mu} t_{\alpha}) - m_f \right] 
\psi_f ,
\eeq
where $G^{\mu\nu}_{\alpha}$ and $A_{\alpha \mu}$ are the color field strength tensor and vector potential, respectively, 
$g$ is the strong coupling constant, $\psi_f$ are the quark fields of different flavors $f$ with masses $m_f$, and $t_{\alpha}$ the generators 
of the color $SU(3)$ group in the fundamental representation. The Lagrangian \eq{qcd} is symmetrical with 
respect to space parity  ${\cal P}$ and charge conjugation parity ${\cal C}$ transformations. 

However, these classical symmetries of QCD become questionable due to the interplay of quantum axial anomaly \cite{anomaly}
and classical topologically non-trivial solutions -- the instantons \cite{Belavin:1975fg}. The axial anomaly arises due to the fact that 
the renormalization of the theory \eq{qcd} cannot be performed in a chirally invariant way. As a result the flavor-singlet axial 
current $J_{\mu 5} = \bar{\psi}_f \gamma_{\mu} \gamma_5 \psi_f$ is no longer conserved even in the 
$m \to 0$ limit:
\beq \label{axanom}
 \partial^{\mu}  J_{\mu 5} = 2 m_f i \bar{\psi}_f \gamma_5 \psi_f - {N_f g^2 \over 16 \pi^2} G^{\mu\nu}_{\alpha} \tilde{G}_{\alpha \mu\nu}.
 \eeq
 where $\tilde{G}_{\alpha \mu\nu} = {1 \over 2} \epsilon_{\mu\nu\rho\sigma} G^{\alpha \rho\sigma}$. The 
 last term in \eq{axanom} is seemingly irrelevant since it can be written down as a full divergence, 
\beq
 G^{\mu\nu}_{\alpha}\tilde{G}_{\alpha \mu\nu} = \partial_{\mu} K^{\mu},
\eeq
of the (gauge-dependent) Chern-Simons current  
\beq\label{topdiv}
K^{\mu} = \epsilon^{\mu\nu\rho\sigma} A_{\alpha \nu} \left[G_{\alpha \rho\sigma} - {g \over 3} f_{\alpha\beta\gamma} 
A_{\beta \rho} A_{\gamma \sigma} \right]. 
\eeq

However this conclusion is premature due to the existence of instantons which 
induce a change in the value of the chiral charge $Q_5 = \int d^3 x K^0$ associated with the topological current 
between $t = - \infty$ and $t= + \infty$: 
\beq 
\nu = \int_{- \infty}^{+ \infty} dt {d Q_5 \over dt} = 2 N_f q[G],
\eeq
where 
\beq\label{topcharge}
q[G] = {\frac{g^2}{32 \pi^2}} \int d^4x G^{\mu\nu}_{\alpha} \tilde{G}_{\alpha \mu\nu}
\eeq
 is the topological charge; 
for a one-instanton solution, $q = +1$. 

In the presence of degenerate topological vacuum sectors, an expectation value 
of an observable ${\cal O}$ has to be evaluated by first computing an average 
\beq
\langle {\cal O} \rangle = \int_q D[\psi] D[\bar{\psi}] D[A] \exp(i S_{QCD}) {\cal O}(\psi, \bar{\psi}, A)
\eeq
over a sector with a fixed topological charge 
$q$, and then by summing over all sectors with the weight $f(q)$ \cite{Weinberg}.  The additivity constraint 
\beq
f(q_1 + q_2) = f(q_1) f(q_2)
\eeq
 restricts the weight to the form 
 \beq
 f(q) = \exp(i \theta q), 
 \eeq
 where $\theta$ is a free parameter. 
Recalling an explicit expression \eq{topcharge} for $q[G]$ one can see that this procedure is equivalent to adding to the QCD Lagrangian  \eq{qcd} $S_{QCD} = \int d^4x {\cal L}_{QCD}$ a new term
\beq\label{theta}
{\cal L}_{\theta} = - {\theta \over 32 \pi^2}  g^2 G^{\mu\nu}_{\alpha} \tilde{G}_{\alpha \mu\nu}.
\eeq
Unless $\theta$ is identically equal to zero, ${\cal P}$ and ${\cal CP}$ invariances of QCD are lost!

Parity violation in strong interactions has been never detected, and stringent limits 
on the value of $\cal{CP}$ violating phase $\theta < 3\times 10^{-10}$ follow from the experimental bounds on the electric dipole moment of the neutron \cite{Baker:2006ts}. Perhaps the most appealing resolution of this puzzle is based on promoting the $\theta$ parameter to a dynamical axion field \cite{Wilczek:1977pj,Weinberg:1977ma} emerging as a Nambu-Goldstone boson of an  additional chiral symmetry \cite{Peccei:1977hh}. We will see that even though in the physical vacuum $\theta$ is equal to zero, the fluctuations of topological charge have an observable effect on QCD phenomena, and the concept of axions allows to formulate an appropriate effective theory.

 \section{Chern-Simons diffusion}

At finite 
 temperature, the transitions between the vacuum states with different topological numbers can occur not only through quantum tunneling, but can also be induced by a classical thermal activation process, through a "sphaleron" \cite{Klinkhamer:1984di}. In electroweak theory sphaleron transitions cause the baryon number violation and may be responsible for at least a part of the observed baryon asymmetry in the Universe \cite{Kuzmin:1985mm}; for a review, see \cite{Rubakov:1996vz}. Sphalerons are also expected to play a r\^{o}le in QCD plasma \cite{McLerran:1990de} where they induce the quark chirality  non-conservation. Unlike for the instanton transitions, the rate of the sphaleron transitions $\Gamma$ is not exponentially suppressed at weak coupling $g$, and in Yang-Mills theory with $N$ colors is proportional to \cite{Arnold:1996dy,Huet:1996sh,Bodeker:1998hm} 
\beq\label{weak}
 \Gamma = const \times (g^2 N)^5 \ln(1/g^2 N) \ T^4
\eeq
  (with a numerically large pre-factor \cite{Moore:1997sn}). Sphalerons describe a random walk in the topological number; in a volume $V$ and after a (sufficiently long) time period $T$ we get the topological number $\langle \nu^2 \rangle =  \Gamma\ V\ T$. The diffusion of topological charge of course is expected to occur not only at weak coupling; while we cannot compute the corresponding rate analytically in QCD, lattice calculations indicate a large rate at experimentally accessible temperatures of $T = 200 \div 300$ MeV \cite{Moore:1997sn}. 
  
A valuable insight is offered also by the ${\cal N} =4$ maximally super-symmetric Yang-Mills theory (N=4 SYM) where the topological charge diffusion rate can be evaluated in the strong coupling limit through the AdS/CFT correspondence \cite{AdS-CFT,AdS-CFT1,AdS-CFT2}, with the following result \cite{Son:2002sd}: 
 \beq
 \Gamma = (g^2 N)^2/(256\pi^3)\ T^4,
 \eeq
which shows that the topological transitions become more frequent at strong coupling, even though the dependence on the coupling is weaker than suggested by \eq{weak}. Note that the large $N$ behavior is the same in the weak and strong coupling limits ($\sim N^0$). 

An explicit example of a topological solution in weakly coupled $AdS_5 \times S_5$ supergravity (that is dual to the large $N$, strongly coupled N=4 SYM) is provided by the D-instanton \cite{Gibbons:1995vg,Green:1997tv} that can be obtained from solitonic D-brane solutions by wrapping them around an appropriate compact manifold. D-instanton can be viewed in the string frame  as an Einstein-Rosen wormhole connecting two asymptotically Euclidean regions of space-time, with the Ramond-Ramond (R$\otimes$R) charge flowing down the throat of the wormhole \cite{Gibbons:1995vg}. It describes a violation of the conservation of a global charge in physical processes \cite{Gibbons:1995vg}. Since in the AdS/CFT dictionary the R$\otimes$R scalar (the axion) of supergravity is dual to the $\theta$ angle of N=4 SYM field theory \cite{Banks:1998nr}, this flow of R$\otimes$R charge down the throat of the wormhole describes the change of topological charge in the field theory description. D-instantons have recently been considered \cite{KL2009} as a source of multiparticle production in high energy collisions in strongly coupled N=4 SYM.

\section{Topologically induced effects in electrodynamics: \\ Maxwell-Chern-Simons theory}
 
\subsection{The Lagrangian} 
 
\begin{wrapfigure}{r}{0.3 \textwidth}
\noindent
%\vspace{-0.2cm}
%\begin{minipage}[b]{.5\linewidth}
\includegraphics[width=3cm]{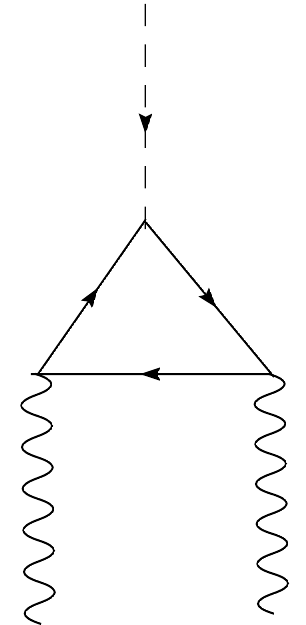}
%\end{minipage}\hfill
%\begin{minipage}[b]{.5\linewidth}
%\includegraphics[width=7.5cm]{fig_monopole}
%\end{minipage}\hfill

\vspace{0.4cm}
\caption{
The quarks  couple the topological Chern-Simons current to the electromagnetic field through the axial anomaly.  
}
\vspace{0.5cm}
\label{fig:axial}
\end{wrapfigure}

Let us begin by coupling the theory \eq{qcd} to electromagnetism; the resulting theory possesses $SU(3) \times U(1)$ gauge symmetry:
$$
{\cal L}_{\rm QCD+QED} =  -{1 \over 4} G^{\mu\nu}_{\alpha}G_{\alpha \mu\nu}  + \sum_f \bar{\psi}_f \left[ i \gamma^{\mu} 
(\partial_{\mu} - i g A_{\alpha \mu} t_{\alpha} -  i q_f A_{\mu}) -  m_f \right] 
\psi_f  - 
$$
\beq\label{qcd+qed}
- {\theta \over 32 \pi^2}  g^2 G^{\mu\nu}_{\alpha} \tilde{G}_{\alpha \mu\nu} - \frac{1}{4}F^{\mu\nu}F_{\mu\nu},
\eeq 
where $A_{\mu}$ and $F_{\mu\nu}$ are the electromagnetic vector potential and the corresponding field strength tensor, and $q_f$ are the electric charges of the quarks.   
\vskip0.3cm
Let us discuss the electromagnetic sector of the theory  \eq{qcd+qed}. Electromagnetic fields will couple to the electromagnetic currents $J_\mu = \sum_f  q_f \bar{\psi}_f \gamma_\mu \psi_f$.  
%Suppose we are interested in an effective theory emerging from \eq{qcd+qed} at large distances $x %\gg \Lambda^{-1}_{\rm QCD}$. In that case the color degrees of freedom decouple, so the sources %will be purely electromagnetic and will be  described by  the  electromagnetic current $J_{\mu}$. 
In addition, the term \eq{theta} will induce through the quark loop the coupling of $F \tilde{F}$ to the QCD topological charge (see Fig.\ref{fig:axial}). We will introduce an effective pseudo-scalar field $\theta = \theta(\vec x, t)$ (playing the r\^{o}le of the axion field) and write down the resulting effective Lagrangian as
\beq\label{MCS}
{\cal L}_{\rm MCS} = - \frac{1}{4}F^{\mu\nu}F_{\mu\nu} - A_\mu J^\mu - \frac{c}{4}\ \theta \tilde{F^{\mu\nu}}F_{\mu\nu},
\eeq
where 
\beq\label{coef}
c = \sum_f q_f^2 e^2 / (2\pi^2). 
\eeq
%{\bf check the coefficient and sign of $A_\mu J^\mu$}  

This is the Lagrangian of Maxwell-Chern-Simons, or axion, electrodynamics that has been introduced previously in \cite{Wilczek:1987mv,Carroll:1989vb,Sikivie:1984yz}. 
If $\theta$ is a constant, then the last term in \eq{MCS} represents a full divergence 
\beq\label{an_ab}
\tilde{F^{\mu\nu}} F_{\mu\nu} = \partial_\mu J_{CS}^\mu
\eeq
of the Chern-Simons current 
\beq\label{topdiv1}
J_{CS}^{\mu} = \epsilon^{\mu\nu\rho\sigma} A_{\nu} F_{\rho\sigma}, 
\eeq
which is the Abelian counterpart of \eq{topdiv}.
Being a full divergence, this term 
does not affect the equations of motion and thus does not affect the electrodynamics of charges.

The situation is different if the field $\theta = \theta(\vec x, t)$ varies in space-time.      
Indeed, in this case we have
\beq
\theta \tilde{F^{\mu\nu}} F_{\mu\nu} = \theta \partial_\mu J_{CS}^\mu = \partial_\mu\left[\theta J_{CS}^{\mu}\right] - \partial_\mu \theta  J_{CS}^{\mu}.
\eeq
The first term on r.h.s. is again a full derivative and can be omitted; introducing notation
\beq
P_\mu = \partial_\mu \theta = ( M, \vec P )
\eeq
we can re-write the Lagrangian \eq{MCS} in the following form:
\beq\label{CS}
 {\cal L}_{\rm MCS} = - \frac{1}{4}F^{\mu\nu}F_{\mu\nu} - A_\mu J^\mu + \frac{c}{4} \ P_\mu J^\mu_{CS}.
\eeq
Since $\theta$ is a pseudo-scalar field, $P_\mu$ is a pseudo-vector; as is clear from   \eq{CS}, 
it plays a r\^{o}le of the potential coupling to the Chern-Simons current \eq{topdiv1}. However, unlike the vector potential $A_\mu$, $P_\mu$ is not a dynamical variable and is a pseudo-vector that is fixed by the dynamics of chiral charge -- in our case, determined by the fluctuations of topological charge in QCD.
\vskip0.3cm
In $(3+1)$ space-time dimensions, the pseudo-vector $P_\mu$ selects a direction in space-time and thus breaks the Lorentz and rotational invariance \cite{Carroll:1989vb}: the temporal component $M$ breaks the invariance w.r.t. Lorentz boosts, while the spatial component $\vec P$ picks a certain direction in space. On the other hand, in $(2 + 1)$ dimensions there is no need for the spatial component $\vec P$ since the Chern-Simons current \eq{topdiv1} in this case reduces to the pseudo-scalar quantity $\epsilon^{\nu\rho\sigma} A_{\nu} F_{\rho\sigma}$, so the last term in \eq{CS} takes the form
\beq\label{2+1}
\Delta {\cal L} = c\ M \epsilon^{\nu\rho\sigma} A_{\nu} F_{\rho\sigma}.
\eeq
This term is Lorentz-invariant although it still breaks parity. 
In other words, in $(2+1)$ dimensions the vector $\vec P$ can be chosen as a 3-vector pointing in the direction of an "extra dimension" orthogonal to the plane of the two spatial dimensions. When added to the Maxwell action, \eq{2+1} generates a mass of the photon which thus becomes "topologically massive". This illustrates an important difference between the r\^{o}les played by Chern-Simons term in even and odd number of space-time dimensions.   
   
\subsection{Maxwell-Chern-Simons equations}   
   
Let us write down the Euler-Lagrange equations of motion that follow from the Lagrangian \eq{CS},\eq{topdiv1}  
(Maxwell-Chern-Simons equations):
\beq
\partial_\mu F^{\mu\nu} = J^\nu - P_\mu \tilde{F}^{\mu\nu}.
\eeq
The first pair of Maxwell equations (which is a consequence of the fact that the fields are expressed through the vector potential) is not modified:
\beq
\partial_\mu \tilde{F}^{\mu\nu} = J^\nu.
\eeq   
It is convenient to write down these equations also in terms of the electric $\vec E$ and magnetic $\vec B$ fields:
\beq\label{MCS1}
\vec{\nabla}\times \vec{B} - \frac{\partial \vec{E}}{\partial t} = \vec J + c \left(M \vec{B} - \vec{P} \times \vec{E}\right), 
\eeq
\beq\label{MCS2}
\vec{\nabla}\cdot \vec{E} = \rho + c \vec{P} \cdot \vec{B},
\eeq
\beq\label{MCS3}
\vec{\nabla}\times \vec{E} +  \frac{\partial \vec{B}}{\partial t} = 0,
\eeq
\beq\label{MCS4}
\vec{\nabla}\cdot \vec{B} = 0,
\eeq
where $(\rho, \vec J)$ are the electric charge and current densities.
One can see that the presence of Chern-Simons term leads to essential modifications of the Maxwell theory. We will now examine a few examples (some old and some new) illustrating the dynamics contained in Eqs\eq{MCS1}--\eq{MCS4}.

\subsection{The Witten effect}

\begin{figure}[htb]
\noindent
\vspace{-0.2cm}
%\begin{minipage}[b]{.5\linewidth}
\includegraphics[width=12cm]{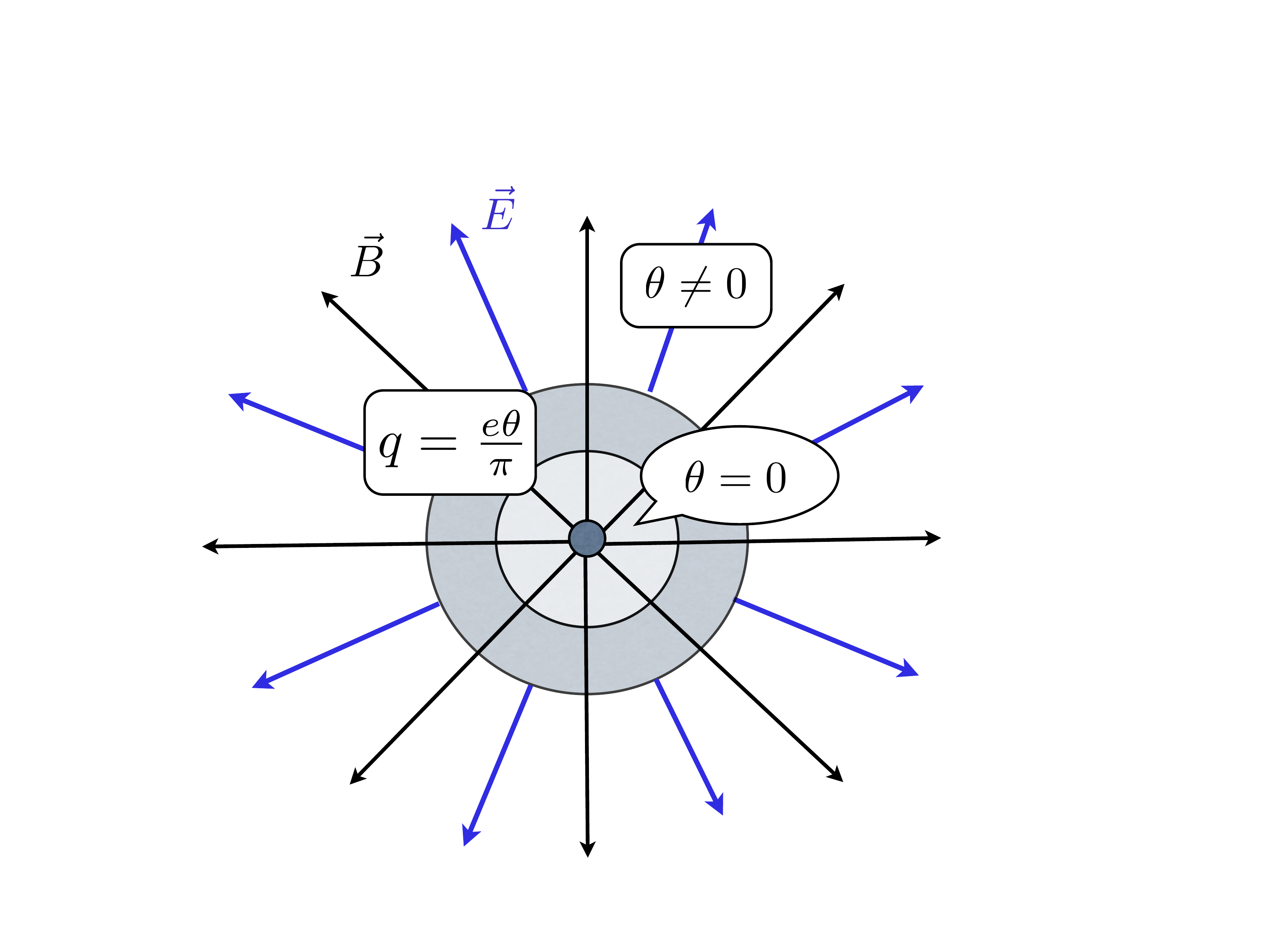}
%\end{minipage}\hfill
%\begin{minipage}[b]{.5\linewidth}
%\includegraphics[width=7.5cm]{fig_monopole}
%\end{minipage}\hfill

\vspace{0.4cm}
\caption{
Magnetic monopole at finite $\theta$ angle acquires an electric charge $\sim e\theta/\pi$ that is localized on the domain wall where 
the value of $\theta$ changes from zero in the core of the monopole to some value $\theta \neq 0$ away from the monopole (the domain wall is shown by the gray ring) -- the Witten effect \cite{Witten:1979ey}.   
}
%\vspace{-0.5cm}
\label{fig:eos}
\end{figure}

Our first example will be the celebrated Witten effect 
\cite{Witten:1979ey}: magnetic monopoles at finite $\theta$ angle acquire electric charge and become "dyons". 
Our treatment of this effect will follow Wilczek \cite{Wilczek:1987mv}. Consider a magnetic monopole in the presence of a finite $\theta$ angle. In the core of the monopole $\theta=0$, and away from the monopole $\theta$ acquires a finite non-zero value -- therefore within a finite domain wall we have a non-zero $\vec P = \vec{\nabla} \theta$ pointing radially outwards from the monopole (see Fig.\ref{fig:eos}). According to \eq{MCS2}, the domain wall thus acquires a non-zero charge density $c  \vec{\nabla} \theta \cdot \vec{B}$. An integral along $\vec P$ (across the domain wall) yields $\int dl\ \partial \theta / \partial l = \theta$, and the integral over all directions of $\vec P$ yields the total magnetic flux $\Phi$. By Gauss theorem, the flux is equal to the magnetic charge of the monopole $g$, and the total electric charge of the configuration is equal to 
\beq
q = c\ \theta\ g = \frac{e^2}{2\pi^2}\ \theta\ g = \frac{e}{2\pi^2}\ \theta\ (e g) = e\ \frac{\theta}{2 \pi},
\eeq
where we have used an explicit expression \eq{coef} for the coupling constant $c$, as well as the Dirac condition $g e = 4 \pi \times {\rm integer}$.

\section{Chiral magnetic effect}

\subsection{Charge separation}

\begin{figure}[htb]
\noindent
\vspace{-2cm}
%\begin{minipage}[b]{.5\linewidth}
\includegraphics[width=14cm]{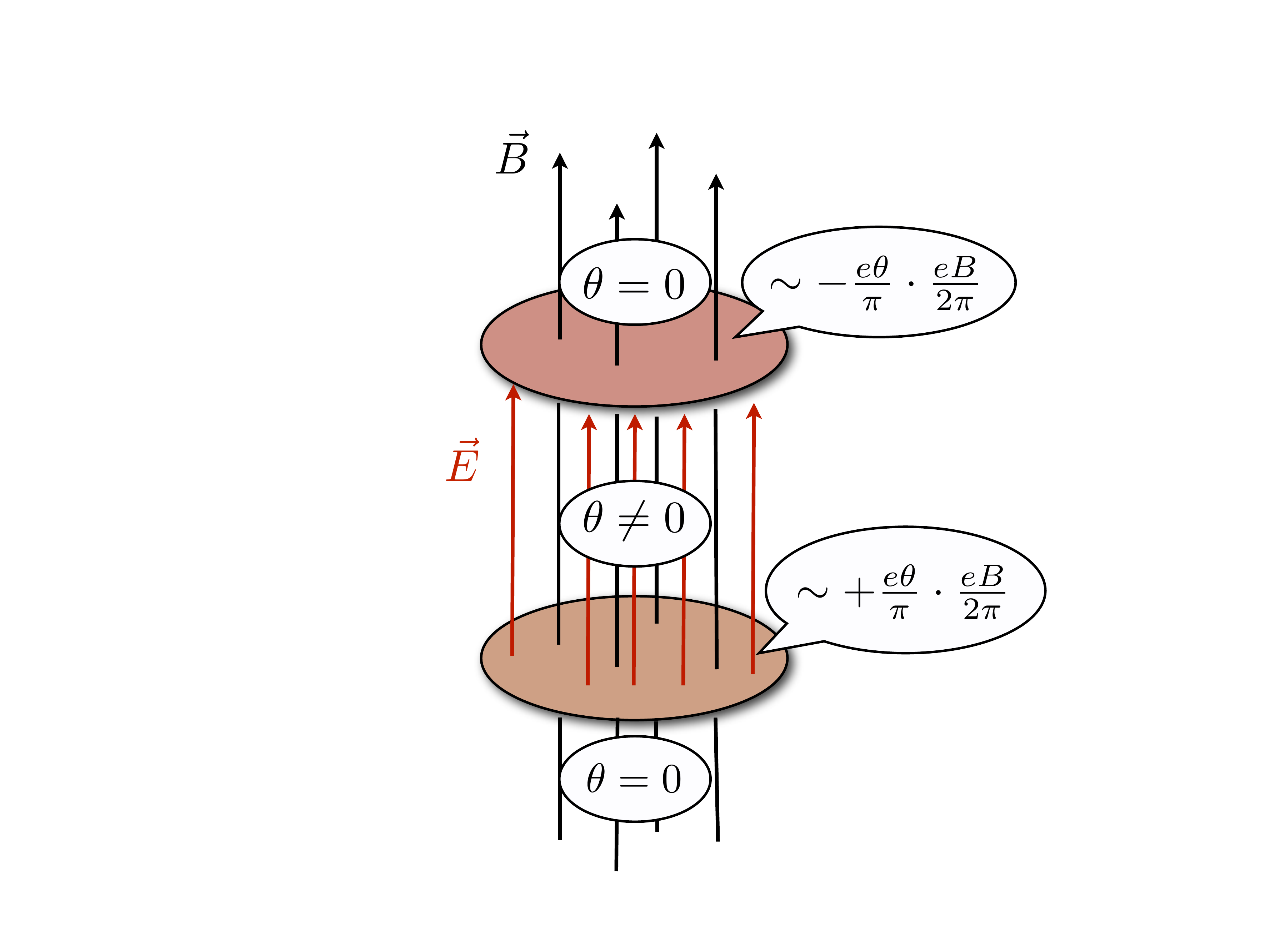}
%\end{minipage}\hfill
%\begin{minipage}[b]{.5\linewidth}
%\includegraphics[width=7.5cm]{fig_monopole}
%\end{minipage}\hfill

\vspace{1cm}
\caption{
Charge separation effect -- domain walls that separate the region of $\theta \neq 0$ from the outside vacuum with $\theta = 0$ become charged in the presence of an external magnetic field, with the surface charge density $\sim e \theta/\pi \cdot eB/2\pi$.
This induces an electric dipole moment signaling ${\cal P}$ and ${\cal CP}$ violation. 
 }
%\vspace{-0.5cm}
\label{chargesep}
\end{figure}

Consider now a configuration shown on Fig.\ref{chargesep} where an external magnetic field $\vec B$ pierces a domain with $\theta \neq 0$ inside;  outside $\theta=0$. Let us assume first that the field $\theta$ is static, $\dot\theta = 0$. Assuming that the field $\vec B$ is perpendicular to the domain wall, 
we find from \eq{MCS2} that the upper domain wall acquires the charge density per unit area $S$ of  \cite{Kharzeev:2007tn}
\beq
\left(\frac{Q}{S}\right)_{up}  = +\ c\ \theta B
\eeq
while the lower domain wall acquires the same in magnitude but opposite in sign charge density
\beq
\left(\frac{Q}{S}\right)_{down}  = -\ c\ \theta B
\eeq 
Assuming that the domain walls are thin compared to the distance $L$ between them, we find that 
the system possesses an electric dipole moment
\beq\label{eldip}
d_e = c\ \theta\ (B \cdot S)\ L = \sum_f q_f^2 \ \left(e\ \frac{\theta}{\pi}\right)\ \left(\frac{eB\cdot S}{2\pi}\right)\ L;
\eeq
in what follows we will for the brevity of notations put $\sum_f q_f^2 = 1$; it is easy to restore this factor in front of $e^2$ when needed.
%\vspace{-0.3cm}
 
Static electric dipole moment is a signature of ${\cal P}$, ${\cal T}$ and ${\cal CP}$ violation (we assume that $\cal{CPT}$ invariance holds). The spatial separation of charge will induce the corresponding electric field $\vec E = c\ \theta\ \vec B$. The mixing of pseudo-vector magnetic field $\vec B$ and the vector electric field $\vec E$ signals violation of ${\cal P}$, ${\cal T}$ and ${\cal CP}$ invariances.

The formula \eq{eldip} allows a simple interpretation: since $eB/2\pi$ is the transverse density of Landau levels of charged fermions in magnetic field $B$, the floor of the quantity $eB\cdot S/2\pi$ (i.e. the largest integer that is smaller than $eB\cdot S/2\pi$) is an integer number of fermions localized on the domain wall. Each fermion species contributes independently to this number as reflected by the factor $N_f$. Again we see that the electric dipole moment \eq{eldip} arises from the electric 
charge $q \sim e \theta/\pi$ that is induced on the domain walls due to the gradient of the pseudo-scalar field $\theta$.  

If the domain is due to the fluctuation 
of topological charge in QCD vacuum, its size is on the order of QCD scale, $L \sim \Lambda_{\rm QCD}^{-1}$, $S \sim \Lambda_{\rm QCD}^{-2}$. This means that to observe an electric dipole moment
in experiment we need an extremely strong magnetic field $eB \sim   \Lambda_{\rm QCD}^{2}$. Fortunately, such fields exist during the early moments of a relativistic heavy ion collision \cite{Kharzeev:2007jp}. Here we have assumed that the domain is static; this approximation requires the characteristic time of topological charge fluctuation $\tau \sim 1/\dot{\theta}$ be large on the time scale at which the magnetic field $B$ varies. This assumption is only marginally satisfied in heavy ion collisions, and so we now need to consider also the case of $\dot\theta \neq 0$.
  
\subsection{The chiral induction}
 
\begin{figure}[htb]
\noindent
\vspace{-1cm}
%\begin{minipage}[b]{.5\linewidth}
\includegraphics[width=14cm]{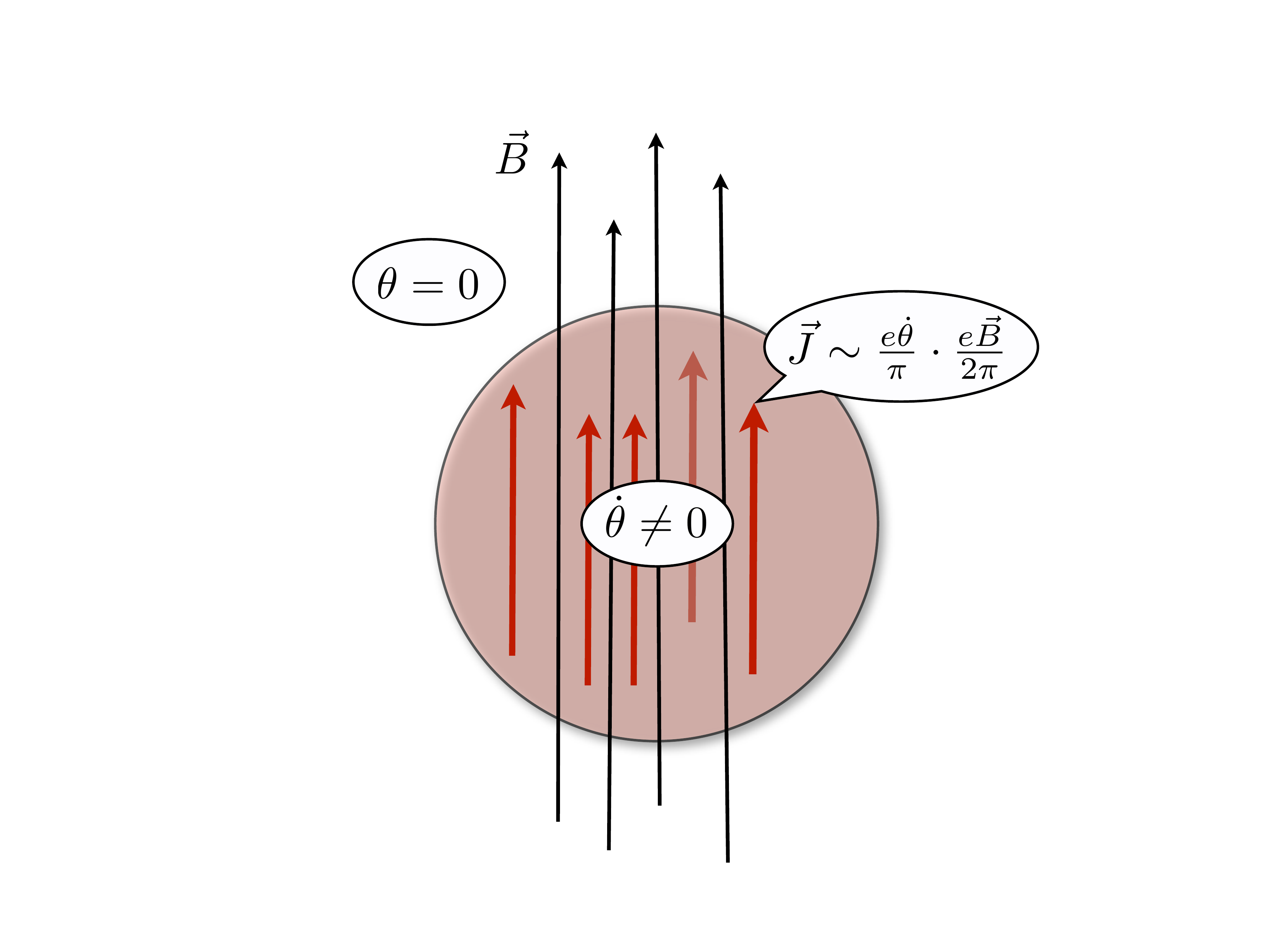}
%\end{minipage}\hfill
%\begin{minipage}[b]{.5\linewidth}
%\includegraphics[width=7.5cm]{fig_monopole}
%\end{minipage}\hfill

\vspace{0.4cm}
\caption{
The chiral magnetic effect -- inside a domain with $\dot\theta \neq 0$ an external magnetic field induces an electric current 
$\vec J \sim e \dot\theta/\pi \cdot e \vec B/2\pi$. $\dot \theta \neq 0$ indicates the change of the chiral charge inducing an asymmetry between the left-- and right-- handed fermions inside the domain. Note that the current $\vec J \sim \vec B$ is absent in Maxwell electrodynamics. 
 }
%\vspace{-0.5cm}
\label{fig:chimagef}
\end{figure}

 Consider now the domain where $| \vec{P} | \ll \dot{\theta}$, i.e. the spatial dependence of $\theta(t, \vec x)$ is much slower than the dependence on time \cite{Kharzeev:2007jp}, see Fig. \ref{fig:chimagef}. Again, we will expose the domain to an external magnetic field $\vec B$ with $\vec{\nabla}\times \vec{B} = 0$, and assume that no external electric field is present.  In this case we immediately get from \eq{MCS1} that there is an induced current \cite{Fukushima:2008xe}
 \beq\label{chimag}
 \vec{J} = - c\ M\ \vec{B} = - \frac{e^2}{2 \pi^2}\ \dot\theta \vec{B}.
 \eeq
Note that this current directed along the magnetic field $\vec B$ represents a ${\cal P}-$, ${\cal T}-$ and ${\cal CP}-$odd phenomenon: 
the electric current is a vector while the magnetic field is a pseudo-vector. This current is 
 of course absent in the "ordinary" Maxwell equations that respect ${\cal P}$ and ${\cal CP}$ invariances. Integrating the current density over time (assuming that the field $\vec B$ is static) we find that when $\theta$ changes from zero to some $\theta \neq 0$, this results in a separation of charge and the electric dipole moment \eq{eldip}.

%%%%%%%%%%%%%%%%%
 
 Let us discuss the meaning of formula \eq{chimag} in more detail. To do this, let us consider the work done by the electric current; to obtain the work per unit time -- the power $P$ -- we multiply both sides of \eq{chimag} by 
 the (static) electric field $\vec{E}$ and integrate them over the volume (as before, we assume that $\theta$ does not depend on spatial coordinates):
 \beq\label{work}
 P = \int d^3 x\ \vec{J}\cdot \vec{E} = - \dot\theta\  \frac{e^2}{2 \pi^2}\  \int d^3 x\ \vec{E}\cdot \vec{B} = - \dot\theta\   \dot{Q}_5,
 \eeq
 where 
 \beq\label{topabel}
 {Q}_5 = \frac{e^2}{2 \pi^2}\  \int dt\ d^3 x\ \vec{E}\cdot \vec{B}
 \eeq
  is the chiral charge. 
 The meaning of the quantity on the r.h.s. of \eq{work} can be revealed with the help of the following well-known quantum-mechanical analogy. The vacuum wave function
\beq
|\theta\rangle = \sum_{Q_5} \exp(i\ \theta\ Q_5)\ |Q_5 \rangle
\eeq 
is analogous to the Bloch wave function of electron in a crystal, with $\theta$ playing the r\^{o}le of electron's quasi-momentum, 
and $Q_5$ -- the r\^{o}le of coordinates of atoms in the crystal. The derivative of the "momentum" $\dot\theta$ thus plays the r\^{o}le of the force that we assume is constant in time and $Q_5$ -- of the dimensionless distance; $\dot{Q_5}$ is thus the velocity. The formula \eq{work} is therefore simply the classical expression
$$
{\rm Power} = {\rm Force} \times {\rm Velocity},
$$
with the force acting along the "extra dimension" of the chiral charge $Q_5$.
\begin{figure}[htb]\label{dirac_cones}
\noindent
\vspace{-1cm}
%\begin{minipage}[b]{.5\linewidth}
\includegraphics[width=14cm]{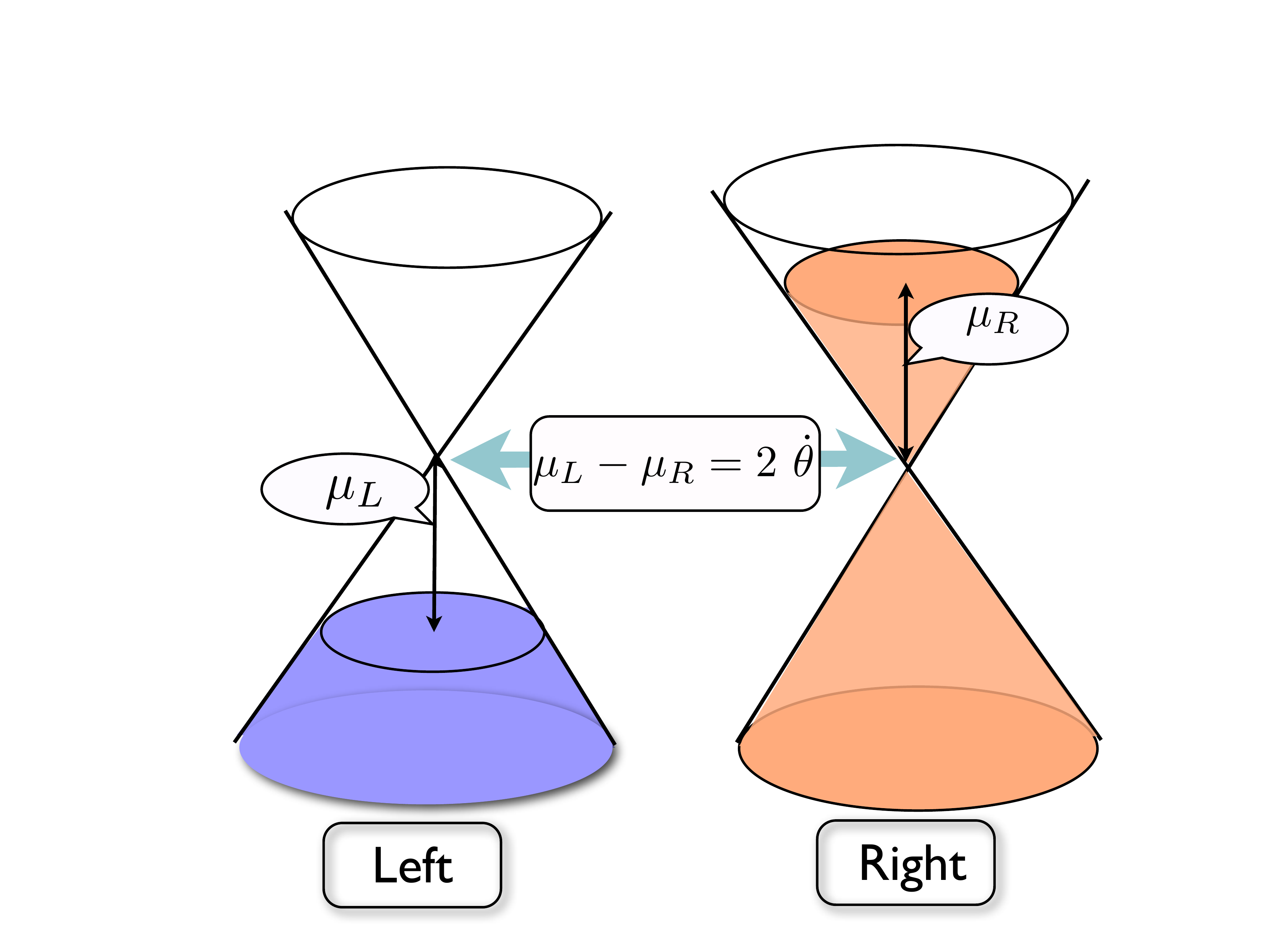}
%\end{minipage}\hfill
%\begin{minipage}[b]{.5\linewidth}
%\includegraphics[width=7.5cm]{fig_monopole}
%\end{minipage}\hfill

\vspace{0.4cm}
\caption{
Dirac cones of the left and right fermions. In the presence of the changing chiral charge there is an asymmetry between 
the Fermi energies of left and right fermions $\mu_L$ and $\mu_R$: $\mu_L - \mu_R = 2 \mu_5 = 2 \dot \theta$. 
 }
%\vspace{-0.5cm}
\label{fig:cones}
\end{figure}
\vskip0.3cm

Let us stress that we have derived  \eq{chimag} from the Maxwell-Chern-Simons equations in the absence of external charges and currents. Therefore, the current \eq{chimag} should arise from the decay of the vacuum state induced by the change of chirality in the presence of magnetic field. Since the vacuum is supposed to have zero energy, what powers the "chiral magnetic" current? To answer this question, consider the physical vacuum of chiral Dirac fermions: all negative energy states of both left and right fermions are occupied, and all positive energy states are empty.  The anomaly relation 
\beq\label{anrel}
Q_5 = N_R - N_L
\eeq
tells us that in the presence of chiral charge there should be an asymmetry between the numbers of left and right fermions, and thus an asymmetry in the corresponding Fermi momenta $p^F_{L,R}$ and Fermi energies $\mu_{L,R} = c\ p^F_{L,R}$, see \fig{dirac_cones}.

Let us compute the densities of left and right fermions in (anti-) parallel electric and magnetic fields; for definiteness let us choose the negative sign for the product $\vec E \cdot \vec B < 0$. The presence of magnetic field $B$ aligns the spins of the positive (negative) fermions in the direction parallel (anti-parallel) to $\vec{B}$. In the electric field $E$ the positive fermions will experience the force $e E$ and will move along $\vec{E}$; therefore their spin will have a positive projection on momentum, and we are dealing with the right fermions. Likewise, the negative fermions will be left-handed. 
After time $t$, the positive (right) fermions will increase their Fermi momentum to $p^F_R = e E t$, and the negative (left) will have their Fermi momentum decreased to $p^F_L = - p^F_R$. The one-dimensional density of states along the axis $z$ that we choose parallel to the direction of fields $\vec E$ and $\vec B$ is given by $dN_R / dz = p^F_R / 2 \pi$. In the transverse direction, the motion of fermions is quantized as they populate Landau levels in the magnetic field. The transverse density of Landau levels is $d^2 N_R/ dx dy = e B/ 2 \pi$. Therefore the density of right fermions increases per unit time as 
\beq
\frac{d^4 N_R}{dt\ dV} = \frac{e^2}{(2 \pi)^2}\ \vec{E} \cdot \vec{B}.
\eeq
The density of left fermions decreases with the same rate, $d^4 N_L / dt\ dV = -  d^4 N_R / dt\ dV$. The rate of chirality $Q_5 = N_R - N_L$ generation is thus
\beq
\frac{d^4 Q_5}{dt\ dV} = \frac{e^2}{2 \pi^2}\ \vec{E} \cdot \vec{B},
\eeq 
which is exactly the $4D$ density of topological charge as given by \eq{topabel}.
\vskip0.3cm
Transferring a fermion from left Fermi surface to the right one costs an amount of energy equal to $\mu_R - \mu_L$, see \fig{dirac_cones}. The power $P$ (work per unit time) of the current \cite{Nielsen:1983rb,Fukushima:2008xe} is thus equal to
\beq 
P = \int d^3 x\ \vec{J}\cdot \vec{E} = (\mu_R - \mu_L)\  \frac{e^2}{(2 \pi)^2}\  \int d^3 x\ \vec{E}\cdot \vec{B}
\eeq
This coincides with \eq{work} if we identify 
\beq
2\ \dot{\theta} = \mu_L - \mu_R \equiv 2 \mu_5,
\eeq
where we have defined a chiral chemical potential $\mu_5$. Again, this matches our interpretation of $\dot{\theta}$ as of the "force" that creates the net momentum $\mu_5/c$ along the "extra dimension" of topological charge $Q_5$. This anomalous mechanism of current generation 
is analogous to the one invoked by Witten \cite{Witten:1984eb} to describe the particle acceleration in cosmic strings. Related condensed matter 
examples have been considered in \cite{Fradkin:1986pq,acf}.

\subsection{Charge separation at finite baryon density and vorticity: \\ the chiral vortical effect}
 
 At finite baryon density, the charge separation and the chiral induction can occur even without an external magnetic field if the angular momentum is present \cite{Kharzeev:2007tn,Kharzeev:2004ey}.  Indeed, let us introduce a matter velocity field $V_{\mu}$; then at finite baryon density $\mu$ Eq. \eq{CS} acquires the following additional term: 
 \beq 
\label{1}
{\cal L}_B = 
- N_c \frac{e \mu}{4 \pi^2}\cdot    \eps\dm \theta\ (\d_{\lambda}V_{\nu} ) A_{\sigma};
 \eeq
this term has been used in the studies of the axial current in cold dense quark matter \cite{SZ,MZ}. It is easy to see that it also 
 induces the electric charge density on the topological domain walls  \cite{Kharzeev:2007tn}:
\beq
\label{2}
\rho_B =\frac{\delta  L_B}{\delta A_0}=
N_c \frac{e \mu }{4 \pi^2}\cdot    \epsilon^{ijk}\d_i \theta\ (\d_{j}V_{k} ) =
  N_c \frac{e \mu}{2 \pi^2}\cdot      \left(\vec{\nabla}\theta\cdot \vec{\Omega}\right),
 \eeq
as well as an electric current directed along the along the angular velocity $ \vec{\Omega}$ of the rotating system, $ 2\epsilon_{ijk} \Omega_k= (\d_i V_j-  \d_{j}V_{i} )$.  In a recent paper \cite{Son:2009tf} it has been shown how this phenomenon manifests itself in relativistic hydrodynamics. 
Since the plasma produced in off-central heavy ion collisions should possess vorticity, this "chiral vortical effect" may be an important source of charge separation, especially at relatively low energies when the net baryon density of the produced matter is significant.

\section{Experimental observations}

\begin{figure}[h]\label{simulations}
\noindent
\vspace{-1cm}
%\begin{minipage}[b]{.5\linewidth}
\includegraphics[width=10cm]{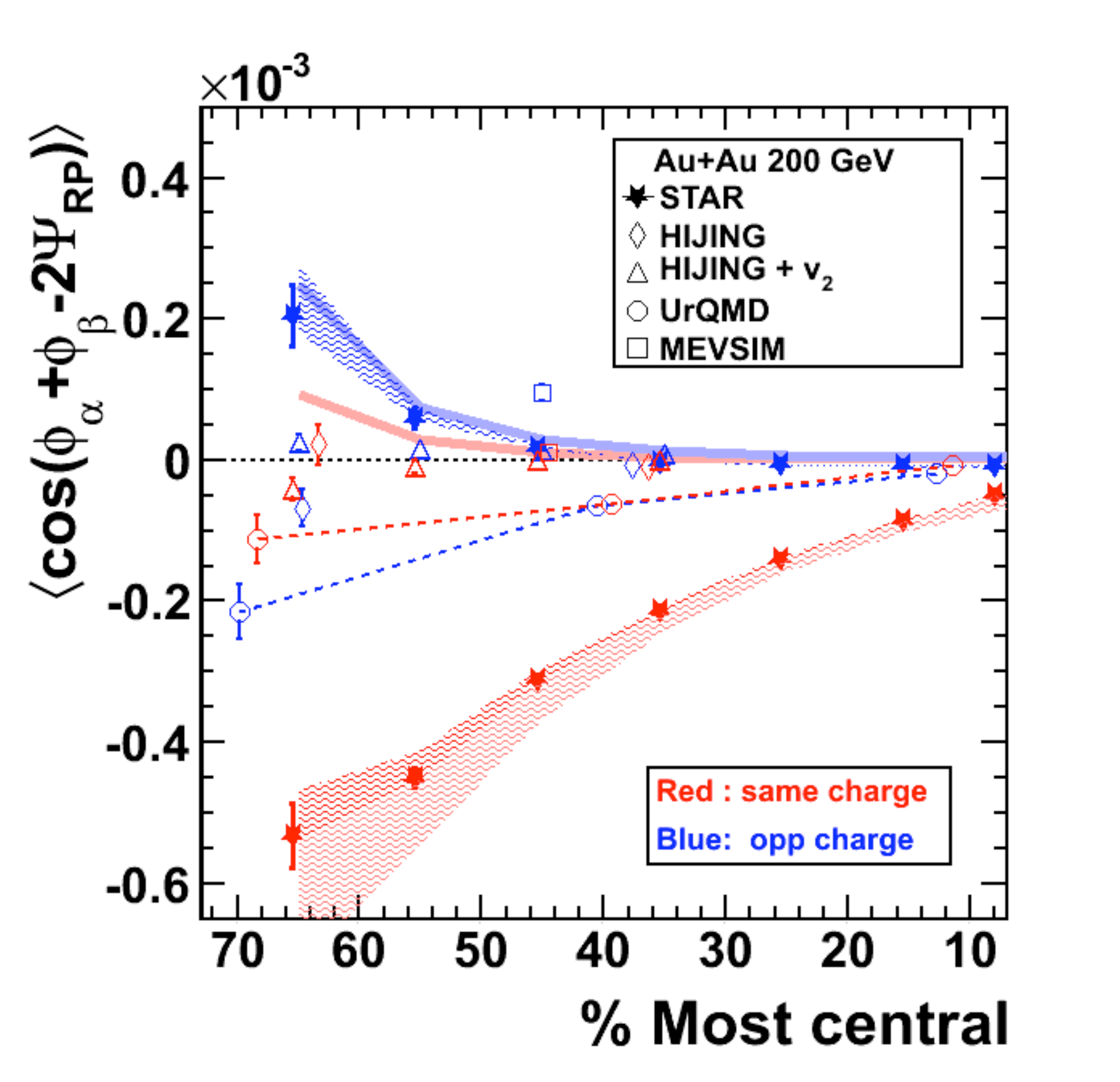}
%\end{minipage}\hfill
%\begin{minipage}[b]{.5\linewidth}
%\includegraphics[width=7.5cm]{fig_monopole}
%\end{minipage}\hfill

\vspace{0.4cm}
\caption{
The STAR Collaboration result on the charge-dependent azimuthal correlations in Au-Au collisions at $\sqrt{s} = 200$ GeV per nucleon pair at RHIC; from \cite{Abelev:2009uh,Abelev:2009tx}. Horizontal axis is the centrality of the collision (the fraction of the inclusive inelastic cross section); 
the average impact parameter decreases towards the right, as does the magnetic field.  Also shown are the predictions of various Monte-Carlo models of heavy ion collisions. 
 }
%\vspace{-0.5cm}
%\label{fig:cones}
\end{figure}
\vskip0.5cm

As we discussed above, the fluctuations of topological charge in the presence of strong magnetic field should induce the separation of electric charges. 
In heavy ion collisions, the magnetic fields of required strength are produced naturally by the electrically charged ions in the initial state, spectators in the final state, and due to the electric charge asymmetry in the distributions of the produced hadrons \cite{Kharzeev:2007jp,Skokov:2009qp}.  The produced magnetic fields are oriented perpendicular to the reaction plane 
(along the system's angular momentum); therefore the chiral magnetic effect under discussion should result in the separation of the electric charge with respect to the reaction plane \cite{Kharzeev:2004ey}. 

The azimuthal distribution of charged hadrons produced in heavy ion collisions can be expanded in Fourier harmonics  in the following way: 
\beq\label{flow}
\frac{d N_{\pm}}{d\phi} \sim 1 + 2 v_1 \cos(\Delta \phi) + 2 v_2 \cos(2 \Delta \phi) + ... + 2 a_\pm \sin(\Delta \phi) + ..., 
\eeq
where $\Delta \phi = \phi - \Psi_{RP}$ is the angle with respect to the reaction plane -- the plane which contains the impact parameter and beam momenta. Note that a typical relativistic heavy ion collision produces several thousand hadrons, so the reaction plane can be reliably identified in each event.  The coefficients $v_1$ and $v_2$ measure the strength of so-called directed and elliptic flow. 
The chiral magnetic effect causes the hadrons with opposite electric charges to be produced preferentially on opposite sides of the reaction plane; 
this corresponds to non-zero coefficients $a_+$ and $a_-$ that should have the opposite signs; this form of the angular distribution follows from the structure of $\cal P$-violating amplitude \cite{Kharzeev:2004ey}. 

The number of charged hadron tracks in the single event (although sufficient to determine the reaction plane) is not large enough to allow a statistically sound extraction of the coefficients $a_\pm$, so one has to sum over many events. However since there is no {\it global} violation of ${\cal P}$ and ${\cal CP}$ invariances in QCD, the sign of the charge asymmetry should fluctuate from event to event and so when averaged over many events, $\left< a_+ \right> = \left< a_+ \right> =0$. The way out of this dilemma has been proposed  by Voloshin \cite{Voloshin:2004vk} who 
suggested to extract the cumulant 
$\left< a_\alpha a_\beta \right>$ by measuring the expectation value of   $\left< \sin(\Delta \phi_\alpha) \sin(\Delta \phi_\beta) \right>$. The variable that 
was proposed in \cite{Voloshin:2004vk}:
\beq\label{variable}
\left< \cos(\phi_\alpha + \phi_\beta - 2 \Psi_{RP}) \right> = \left< \cos \Delta \phi_\alpha \cos \Delta \phi_\beta \right> - \left< \sin \Delta \phi_\alpha \sin \Delta \phi_\beta \right>
\eeq
has an added benefit of not being sensitive to the reaction plane--independent backgrounds that cancel out in \eq{variable}. The variable \eq{variable} can be measured with a very high precision, and is directly sensitive to the parity--odd fluctuations; the price to pay however is that the observable itself is parity--even, and so one has to carefully examine all possible physics backgrounds.

The preliminary results have been reported by STAR Collaboration in \cite{Selyuzhenkov:2005xa,Voloshin:2008jx}. Very recently, the STAR Collaboration has presented a conclusive measurement of \eq{variable} that amounts to the observation of charge-dependent azimuthal asymmetries in heavy ion collisions at RHIC \cite{Abelev:2009uh,Abelev:2009tx}; one of the results is shown in Fig.\ref{simulations}. One can see that the same-charge and opposite-charge cumulants \eq{variable} differ in a way that is very significant statistically. Numerous mundane backgrounds have been examined, and none of them could explain the observed effect so far \cite{Abelev:2009uh,Abelev:2009tx}. The predictions of various Monte Carlo models of heavy ion collisions are also shown in Fig.\ref{simulations}; these models (while successful in reproducing the global features of heavy ion collisions) fail in explaining the observed effect. Needless to say, one has to continue looking for alternative explanations of the STAR result, and to extend the experimental study of charge asymmetries; however in any case the observed effect is clearly novel and very intriguing.

\section{Summary and discussion}

In this paper we discussed how the fluctuations of topological charge in the presence of strong magnetic field induce the separation of electric charge along the axis of magnetic field (the chiral magnetic effect) \cite{Kharzeev:2004ey,Kharzeev:2007tn,Kharzeev:2007jp,Fukushima:2008xe,Kharzeev:2009pj}. 
The evidence for the chiral magnetic effect from lattice QCD has been found recently in \cite{Buividovich:2009wi}. The behavior of the effect at strong coupling in the Sakai-Sugimoto model and related theories has been explored through the AdS/CFT correspondence in \cite{Lifschytz:2009sz,Yee:2009vw,Rebhan:2009vc}. Recently, the STAR Collaboration at RHIC has presented an observation of charge-dependent azimuthal asymmetries \cite{Abelev:2009uh,Abelev:2009tx}.  None of the mundane backgrounds could explain the observed effect so far \cite{Abelev:2009uh,Abelev:2009tx}. It is clear that a dedicated experimental program is necessary to understand fully this intriguing observation. In particular, one could extend the present studies of charge asymmetries to identified hadrons and to look for the possible weakening of the observed effect at low energies where the energy density is under the threshold for forming the QCD plasma.  Both studies have already been planned at RHIC.
\vskip0.3cm
Here we discussed the sphaleron transitions as a way of generating topological charge in the QCD plasma. 
There exists however a number of other possibilities: the metastable parity--odd domains \cite{Kharzeev:1998kz,Kharzeev:1999cz,Halperin:1998gx,Fugleberg:1998kk} that emerge in the large N description of plasma in QCD and related theories; and the dyons that have been identified in the plasma in lattice QCD \cite{Chernodub:2009hc,Bornyakov:2007fm}, to mention a few. Whatever the specific "microscopic" origin of the topological fluctuations, the fact of their existence in the plasma is firmly established by now in numerical lattice simulations. The recent study \cite{Iqbal:2009xz} finds that the topological charge density 
in QCD plasma at temperatures $T_c \leq T \leq 2 T_c$ is correlated over distances of order $1/T$, and that these correlations are much stronger than 
in the vacuum; for a review of the earlier work, see \cite{Vicari:2008jw}. It is important to note that topological fluctuations exist already at the early stages of the collision prior to thermalization, both in the weak \cite{Kharzeev:2000ef,Shuryak:2000df,Nowak:2000de,Kharzeev:2001ev,Lappi:2006fp} and strong \cite{KL2009} coupling descriptions. 

\vskip0.3cm

The mechanism considered here requires a sufficiently large energy density for the topological transitions to turn on, and for the quarks to separate by a distance comparable to the system size -- therefore, there has to be a deconfined phase. In addition the system has to be in a chirally symmetric phase -- in a chirally broken phase, the chiralities of quarks could flip easily causing dissipation of the induced current.  
The experimental and theoretical studies of parity--odd charge asymmetries in heavy ion collisions can thus bring us closer to the detection 
of a change  in the physical properties of the vacuum induced by creating a high energy density over an extended volume \cite{Lee:1974ma}. 

\vskip0.3cm
{\it Note added:} Very recently, Blum and collaborators \cite{Abramczyk:2009gb} have reported on the first study of the chiral magnetic effect in lattice gauge theory with dynamical $2+1$ flavors 
of quarks (in the domain wall formulation).   

  \section*{Acknowledgments}
  I am grateful to K. Fukushima, Yu. Kovchegov, A. Krasnitz, E. Levin, L. McLerran, R. Pisarski, M. Tytgat, R. Venugopalan, H. Warringa, and A. Zhitnitsky for enjoyable collaborations on the topics discussed here. I am indebted to T.D. Lee, J. Sandweiss and S. Voloshin for many 
  enlightening discussions. 
This work was supported by the U.S. Department of Energy under Contract No. DE-AC02-98CH10886. 


\end{document}